\documentclass[12pt]{iopart}
\usepackage{iopams}
\usepackage{graphicx}
\usepackage{dcolumn}
\usepackage{bm}

\begin{document}

\title{Casimir Forces in Multi-Sphere Configurations}

\author{James Babington}
\address{%
Laboratoire de Physique et Mod\'{e}lisation des Milieux
Condens\'{e}s,\\ Universit\'{e} Joseph Fourier and CNRS, Maison des
Magist\`eres, \\ 38042 Grenoble, France.}
\ead{james.babington@grenoble.cnrs.fr}
\author{Stefan Scheel}%
\address{%
Quantum Optics and Laser Science, Blackett Laboratory, \\
Imperial College London, Prince Consort Road, London SW7 2AZ, U.K.}
\date{\today}

\begin{abstract}
We calculate the Casimir force on an isolated dielectric sphere in an ensemble of $N$ spheres due to multiple mutual interactions of the collection of spheres. In particular we consider dielectric spheres immersed in some other background dielectric. As an example, the Casimir force between two and three
spheres at zero and finite temperature is evaluated. For a very large number of spheres, we consider a large-$N$ scaling limit of the Casimir force.
\end{abstract}

\maketitle

\section{\label{sec:INTRODUCTION}Introduction}

Dispersion interactions between macroscopic bodies have been the subject of investigation for a long time, and a variety of experiments have been performed to test the theoretical predictions. Whilst the theory of Casimir forces~\cite{Casimir:1948dh} between bodies in free space is well defined, the issue of choosing the correct stress tensor for bodies immersed in dielectric backgrounds is not a closed discussion~\cite{raabe:013814,pitaevskii:047801,raabe:047802, Brevik2009, pfeifer2007}. Depending on what type of compatibility is required with the microscopic degrees of freedom, one is led to different choices. The classic example of the Minkowski vs. canonical stress tensor (formally the same as the stress tensor in microscopic electrodynamics), or the choice for the momentum density vs. Poynting vector describing momentum density propagation in a dispersive media show the different possibilities on offer.  The canonical stress tensor is the choice compatible with the Lorentz force law~\cite{raabe:013814} that describe the microscopic degrees of freedom. With this choice, the formalism of macroscopic quantum electrodynamics~\cite{scheel-2008-58} can be used to evaluate the two point correlation functions. In this description one is able to maintain the equal time commutation relations for the physical fields and thus employ a sensible quantum theory of light interacting linearly and causally with matter. In the same fashion, many calculations are successfully carried out in the well established Lifshitz framework~\cite{Lifshitz1961,Lifshitz1980,advancescasimir}. It is also not surprising that there should be a large overlap with colloid physics and the corresponding dispersion forces (see for instance~\cite{ninham1976}). Given that Casimir force experiments can now be performed with bodies immersed in dielectric media~\cite{PhysRevA.78.032109,zwol.041605,Phys.Rev.Lett.101}, it is an important objective to develop the predictions based on the canonical stress tensor for future experimental matching. With this testing in mind, we may start to think of new physical probes e.g. many body interactions~\cite{PhysRevLett.92.078301} and critical Casimir effects~\cite{Nature.451.172.2008} as a way of better understanding and testing the underlying processes. This combined effort is necessary in establishing any scheme for the choice of the relevant physical observables.

The necessity of a detailed understanding of the above issues is clearly of particular relevance to nano-scale architectures, where a precise knowledge of these forces (stiction forces) between nano-mechanical objects is important. At such small length scales where bodies are attracting/repelling due to these interactions, one will encounter new frictional forces or many body effects. Atom chip miniaturisation~\cite{Zimmermann} is a good example where one would need a quantitative description. We also see the point here that a detailed understanding of both different geometries and many body effects will be required. 

Clearly the practical need is to better understand the nature of the interactions between macroscopic bodies. As recently pointed out in~\cite{messina-2009-80} one needs to go beyond the simple proximity force approximation (PFA) and correctly include volume (bulk) contributions to the Casimir force (see also~\cite{advancescasimir} for an overview). To this end, Casimir energies have been calculated recently~\cite{emig-2007-99,emig-2008-41,Rahi:2009hm,kenneth:014103,Bulgac:2005ku} between arbitrary compact bodies. Their approach was to evaluate a suitable energy functional by the use of path integral determinants and to make expansions, having evaluated a T-operator (see also~\cite{BalianBloch,BalianDuplantier,renne} for similar early work). By deducing an interaction potential, a force could then be derived by simple differentiation (for an overview, see e.g. Ref.~\cite{Milton:2008st}). Related work on both calculational and matters of principle  can be found in~\cite{Barton1, Lambrecht2006,PhysRevD.73.125018,golestanian-2009,rodriguez-2009}.

In this article we evaluate the Casimir force for a collection of dielectric spheres in a dielectric media using  the framework of macroscopic quantum electrodynamics as applied in Ref.~\cite{raabe:013814} to the planar geometry. We use the canonical stress tensor to calculate the resultant force on one of the spheres in the configuration. A brief comment will be made on how they qualitatively differ from those that arise from the Minkowski stress tensor, though a detailed comparison of quantitative differences will be left to the future (when the appropriate experiments have the necessary sensitivity). The permittivities are considered to be frequency dependent and complex in general. The examples chosen to illustrate the inter-sphere forces will be evaluated in the retarded limit whereby we will use their constant static values. The calculations are carried out in a similar fashion to the Casimir energies cited above, by performing perturbative
expansions of scattering two-point functions in terms of the relevant reflection/scattering coefficients. That these are the relevant perturbation parameters follows directly from energy conservation (i.e. the dielectric spheres are not amplifying bodies) and that they
are effectively the energy dependent coupling constants of the theory. Generally speaking, one needs to find an approximate geometry for which the Helmholtz equation can be solved (i.e. we perform the separation of variables in local coordinate patches and glue them together). Then the eigenfunctions and eigenvalues can be used to evaluate the scattering
correlation functions in one of two ways. One is to use a Lippmann-Schwinger evaluation of the Green's function based on the known background Hamiltonian where the dielectric spheres are perturbation potentials on top of this. The second route which is the one followed here is to apply the boundary conditions at all the interfaces and then deduce the `out' states (the scattered modes together with their outgoing wavefunctions) from the `in' states (the incident driving modes and their incoming wavefunctions). The driving modes, which are the coefficients of the eigenfunction expansion of the electric and magnetic fields (the precise form will be given later on), must be evaluated and indeed these are deduced from the background when no spheres are present. 

In fact, although one usually talks about the separation of the spheres, it is the total path length that is the meaningful distance when talking about the $N$-body force. A further
simplification is possible for a set of terms in the multiple scattering expansion where the number of spheres becomes large, whilst the coupling constants become small. For some fixed separation of the spheres we can deduce an alternative functional form of the force.

The article is organised as follows. In Sec.~\ref{section-force} we
evaluate the scattering two-point functions for the given setup of $N$
distinct spheres as a perturbative series in the Mie scattering
coefficients. This allows us to write the Casimir force as a
multiple scattering expansion. In Sections~\ref{sec:2sphereForce} and~\ref{sec:3sphereForce} we
consider the two-sphere and three sphere configurations both at zero and finite
temperatures. The evaluation performed here is for the retarded limit where we take the static values of the permittivities, as a definite way in which to evaluate frequency integral. In Sec.~\ref{sec:Nsphere} we calculate the
$N$-body force in the limit of a large number of spheres and at weak
coupling. Finally in Sec.~\ref{sec:conclusions} we set out our
conclusions.

\section{The general force expression in an N-sphere system}
\label{section-force}

It is worth remarking on some aspects of the geometry for the simple
two sphere system before considering the general $N$-sphere
system. One would expect the force on one of the two spheres (with separation $r$) can
be evaluated in three distinct perturbative regions i.e. particular distinct separations. In each region the force will have some scaling law:- 
\begin{enumerate}
\item Region 1, where the separation of the spheres is much greater
than the sphere radii. Here an asymptotic multipole expansion is
relevant. The polarizabilities act as the perturbation theory coupling
constants. The leading term for the force goes as $1/r^8$ and the effective geometry is three dimensional. 
\item Region 2, where the separation of the sphere surfaces is small
compared to the radii. The leading term for the force per unit area goes as $1/r^4$ for small parallel elements of area. The effective
geometry becomes planar (one dimensional). This would be the basis for the
proximity force approximation (PFA) where curvature corrections are
then taken into account at close separations. The force then goes as $1/r^3$. 
\item Region 3, where the separation is comparable to the radii. Here
one might think that there exists a third perturbative point (effective geometry) that is somewhere between the forms of regions one and two. The
leading term for the force per unit length may be expected to go as $1/r^6$ with the corresponding two dimensional effective geometry. 
\end{enumerate}
The 'effective geometry' mentioned above is just the space in which the bodies can be moved and that parametrize the interaction. For example the parallel plate geometry has just the one dimension in which they can be moved. By interpolating the known perturbative solutions in regions 1 and 2
one can find an approximate solution that can be used over all
distances. However, this method also requires the use of two different
cutoff separations, where in the lower limit we would encounter an
infinity and an upper cutoff to say what we mean by a 'large' 
separation. Is is also difficult to glue these asymptotic solutions
together in the intermediate region because we need to convert the
small separation pressure into a force. It is not obvious what area
integration to use here and signals the trouble of using the PFA in
high curvature compact geometries. 

\subsection{The scattering mode decomposition}

\begin{figure}[htbp]
\begin{center}
\includegraphics[height=6cm]{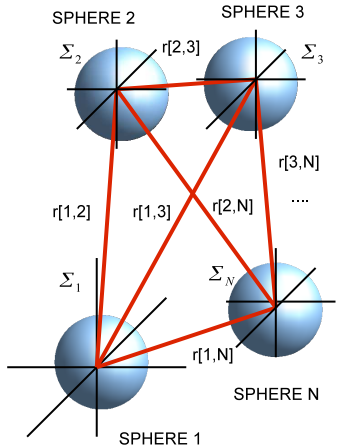}
\caption{The N-Sphere system consists of N dielectric spheres of radii
$R[1],\cdots, R[N]$ each centred on N separate coordinate systems
$\Sigma_1, \cdots ,\Sigma_N$, all contained in a background
dielectric.}\label{fig:NSPHERES} 
\end{center}
\end{figure}

The general strategy in calculating the Casimir force on a particular
body is simply to calculate the scattering 2-point correlation
functions for the physical fields, in our case the electric and
magnetic fields. Given $N$-spheres we are going to calculate the force
on sphere 1 due to the presence of the remaining $(N-1)$ spheres. The
Helmholtz equation for the electric and magnetic fields is solved in
$N$ separate coordinate systems centred on each sphere  in terms of
vector wave functions about each of the separate spherical coordinate
systems origin~\cite{Mackowski06081991}, \cite{Moneda:07}. The
pictorial setup of our system is given in Figure~\ref{fig:NSPHERES}. 

Using the results in~\cite{Mackowski06081991}, one is able to map
solutions about any given sphere (any coordinate system $\Sigma_i$)
into a solution about any other translated origin (a coordinate system
$\Sigma_j$). It is this fact that will enable us to perform the
integral over the volume of the sphere. 

The Casimir force on sphere-1 (with volume $B^2$, the ball which has the two-dimensional sphere as its boundary) due to the effects of the N-sphere
system of differing material properties is given by 
\begin{equation}
\label{eq:stressintegral}
F_j(1|N-1)= \int_{B^2} d^3x \nabla_i T_{ij}(x).
\end{equation}
The stress-energy tensor is given by the standard vacuum expression
(which is consistent with the Lorentz force law~\cite{raabe:013814}) 
\begin{equation}
T_{ij}(x)=\mathbf{E}_i(x)\mathbf{E}_j(x)+\mathbf{B}_i(x)\mathbf{B}_j(x)
-\frac{1}{2}\delta_{ij}(|\mathbf{E}(x)|^2+|\mathbf{B}(x)|^2),
\end{equation}
and it is understood that we are taking the limit
for the initial and final points 
\begin{eqnarray}
\lim_{x_1\rightarrow x_2}\mathbf{E}(x_1)\mathbf{E}(x_2)
&=&\mathbf{E}(x_2)\mathbf{E}(x_2 ),\\
\lim_{x_1\rightarrow x_2}\mathbf{B}(x_1)\mathbf{B}(x_2)
&=&\mathbf{B}(x_2)\mathbf{B}(x_2 ).
\end{eqnarray}
In performing this limit it is first necessary to make subtractions of
the singular part arising from the free Green's tensor. This is
somewhat like a vacuum bubble one encounters in quantum field theory
where we are throwing away an unobservable zero point energy.  We then
need to evaluate the scattering correlation functions (whilst dropping
the direct modes of propagation), viz. 
\begin{equation}
\lim_{y \rightarrow x}\mathbf{E}_i(x)\mathbf{E}_j(y)
=\int_{0}^{\infty}
d\omega d\omega^{\prime}\langle
\mathbf{E}^{out}_{i}(x;\omega)^{\dagger}
\mathbf{E}^{in}_{j}(y;\omega^{\prime})\rangle,
\end{equation}
and similarly for magnetic fields. 

A crucial second step is to realise that the volume integral of the
force requires a careful specification of the coordinate origin in
order to be able to evaluate the derivative. In the multiple
scattering approach used here the coordinate origin used will be the
one centred on the sphere from which the last scattering event took
place, then translated to the origin of the sphere where we are
calculating the force. In terms of the classical scattering Green's
tensor $\mathbf{G}_{ij}$ centred on the sphere where we are
calculating the force, schematically we have 

\begin{equation}
\int_{B^2} d^3x \nabla_i \mathbf{G}_{ij}(\mathbf{x},\mathbf{x})
\approx R \int_{S^2}d \Omega_2 R^2  \nabla_{\mathbf{X}}
\mathbf{G}_{rr}(\mathbf{R},\mathbf{R}) , 
\end{equation}
where $\mathbf{X}$ is the vector connecting the centre of the two
spheres, originating from the last scattering sphere centre and $d\Omega_2$ is the element of solid angle. It should be pointed out here that this approximation, in order to be consistent with the case of a sphere and a dipole, requires the introduction of a normalisation factor (for us it will be $1/4\pi$). This is because the derivative acting on the stress tensor is evaluated at the centre of the sphere, whilst the wavefunctions are evaluated on the spheres surface. In the case of the sphere and dipole, we do not have the integral to do over the $S^2$. Due to the potentially singular nature when shrinking the sphere down to a point, this clearly must be done carefully.

To construct the scattering two point function we write the fields in
a mode decomposition~\cite{Mackowski06081991} of spherical vector wave
functions 
\begin{eqnarray}
 |\mathbf{E}^{in}\rangle =\sum_{i,L,m}p^{i}_{L,m}\mathbf{L}
 |L,m,j_L\rangle+  
 \sum_{i,L,m} q^{i}_{L,m}\left(\frac{1}{k}\nabla\wedge
 \mathbf{L} \right)|L,m,j_L \rangle, \\
 |\mathbf{E}^{out}\rangle  =\sum_{i,L,m}a^{i}_{L,m} \mathbf{L}
 |L,m,h_L^{+} \rangle+ 
 \sum_{i,L,m} b^{i}_{L,m}\left(\frac{1}{k}\nabla\wedge
 \mathbf{L} \right)|L,m,h_L^{+} \rangle, \\
 |\mathbf{E}^{internal}\rangle =\sum_{i,L,m}c^{i}_{L,m} \mathbf{L}
 |L,m,j_L \rangle+ 
\sum_{i,L,m} d^{i}_{L,m}\left(\frac{1}{k}\nabla\wedge
\mathbf{L} \right)|L,m,j_L \rangle, \\
 |\mathbf{B}\rangle = \frac{i}{\omega} \nabla \wedge |\mathbf{E} \rangle .
\end{eqnarray}
(Note that the mode operator coefficients are functions of frequency
which for now has been suppressed). The `in' and `internal' states are
regular at the $i$-sphere origin, whilst the `out' states are outgoing
modes falling off at infinity. They are eigenfunction modes with
respect to the $i$-sphere of the radial eigenfunctions, represented by
spherical Bessel and Hankel functions respectively. The internal
coefficients $(c^{i}_{L,m}, d^{i}_{L,m})$ of a given sphere can be
eliminated by application of the continuity equations across each
sphere. One is then able to determine the scattering coefficients
$(a^{i}_{L,m}, b^{i}_{L,m})$ in terms of the driving modes
$(p^{i}_{L,m}, q^{i}_{L,m})$ by making use of a translation addition
theorem for vector wave functions~\cite{Mackowski06081991}. These are
given by 

\begin{eqnarray}
 \mathbf{L} |i1,L1,m1,j_{L1} \rangle =&\sum_{i2,L2,m2} A^{i1,i2}_{L1,m1;L2,m2}\mathbf{L}|i2,L2,m2,h_{L2}^{+}\rangle \nonumber \\
&+B^{i1,i2}_{L1,m1;L2,m2} \frac{1}{k}\nabla \wedge \mathbf{L}
|i2,L2,m2,h_{L2}^{+}  \rangle, 
\end{eqnarray}
and
\begin{eqnarray}
 \frac{1}{k}\nabla\wedge \mathbf{L} |i1,L1,m1,j_{L1} \rangle 
=&\sum_{i2,L2,m2}  A^{i1,i2}_{L1,m1;L2,m2}\frac{1}{k} \nabla \wedge
\mathbf{L}|i2,L2,m2,h_{L2}^{+} \rangle  \nonumber \\
&+ B^{i1,i2}_{L1,m1;L2,m2}\mathbf{L}|i2,L2,m2,h_{L2}^{+} \rangle .
\end{eqnarray}
Here, $A$ and $B$ are representations of the translation group in the
angular momentum basis. They can be found in~\cite{Mackowski06081991} (see also~\cite{Moneda:07}),
together with the other necessary quantities used here, but in
particular they are functions only of the centre-to-centre sphere
separation $\mathbf{r}[i,j]$ and the background $k$. The matrices $A$
map TE-to-TE and TM-to-TM, whilst the matrices $B$ map TE-to-TM and
vice versa. Using the Mie scattering coefficients $(\alpha_L,\beta_L)$
the scattering modes are given by (where we suppress indices) 
\begin{equation}
(a,b)(\mathbf{1} +T)=(\alpha \cdot p,\beta \cdot q), \label{eq:tmatrix}
\end{equation}
where the T-matrix has the block diagonal form 
\begin{equation}
\mathbf{T}=
\left( 
\begin{array}{cc}
 \alpha \cdot A & \alpha \cdot B \\ 
\beta \cdot A &  \beta \cdot B  \\ 
\end{array}
\right).
\end{equation}
The matrix equation given in~(\ref{eq:tmatrix}) can be inverted and
then perturbatively expanded as a power series in the Mie scattering
coefficients (which play the role of the small coupling constants)
such that the scattered modes are found to be 
\begin{equation}
\label{eq;scatmodes}
|a,b\rangle=\sum^{\infty}_{n=0} \prod_{m=0}^{n}(-\mathbf{T})^m|\alpha
\cdot p,\beta \cdot q\rangle. 
\end{equation}
The T-matrix is then the set of reflection coefficients from each
sphere appropriately translated and the integer $n$ gives the number
of scattering events for the whole set of spheres. This is the  multiple
scattering approach (see for example~\cite{Butler1999}) which allows
for all possible scattering between the spheres. 

In order to calculate the $N$-body Casimir force it is necessary to
consider all possible ways of connecting all of the spheres. We can form a primitive set of diagrams
that can be built on to generate all other diagrams. A suitable set of
definitions are:-  
\begin{itemize}
\item Disconnected diagrams are where some subset of the spheres do
not take part in the multiple $N$-body scattering process. 
\item Simply-connected diagrams are where all the spheres are
connected with no more than two lines attached to any sphere. This
forms a continuous path through all the spheres. 
\item Reducible diagrams are where diagrams can be written as simply
connected diagram convoluted with a lower body diagram. This affords a
description of multiple internal scattering. 
\end{itemize}
The simply-connected diagrams can be seen to be irreducible $N$-body
diagrams. Unless otherwise stated, we will always be calculating the
simply-connected diagrams and corresponds to one particular
perturbation series (as opposed to a perturbation series based on the
sphere radii). 

\subsection{Dielectric backgrounds}

We now specialise to the case of dielectric backgrounds and consider an approximation where
$\beta_L=0$ so that the $b$ modes vanish. This approximation is based on the eventual small argument expansion of the Mie scattering coefficients (and thereby a simple multipole expansion) where the $\alpha_L$ scattering coefficient gives the leading contribution in the frequency integral. What is interesting in this
case is that there are only contributions from the cross terms
involving the  $a$ and $p$ modes, as the $q$ and $a$ cross terms are
zero due to the antisymmetry properties of the translation products
$A^n\cdot B$. From Equation~(\ref{eq;scatmodes})  it is possible to
evaluate the 2-point functions in terms of the driving modes on the
surface of sphere-1. The relevant components (suppressing $SO(3)$
indices) contributing to the $rr$ component of the stress tensor are:-

\begin{eqnarray}
 \int_{S^2}\langle \mathbf{E}^{i1,out}_r(x;\omega)^{\dagger}
\mathbf{E}_r^{i2,in}(x;\omega^{\prime})\rangle|_{x=R[1]}
=& 0,\nonumber \\
 \int_{S^2}\langle \mathbf{B}^{i1,out}_{r}(x;\omega)^{\dagger}
\mathbf{B}^{i2,in}_{r}(x;\omega^{\prime})\rangle |_{x=R[1]}
= & \Im\frac{k^2}{\omega^2}
\frac{(L(L+1))^2}{k^2R[1]^2}j_{L}(k|R_1|)h^{+}_{L}(k|R_1|) \nonumber \\
& \cdot \langle a_{L^\prime}^{i1}(\omega)^{\dagger} \cdot
A^{i1,i2}_{L^{\prime} L}\cdot p^{i2}_{L}(\omega^{\prime}) \rangle,\nonumber \\
 \int_{S^2}\langle \mathbf{E}^{i1,out}(x;\omega)^{\dagger}\cdot
\mathbf{E}^{i2,in}(x;\omega^{\prime})\rangle|_{x=R[1]}
=&\Im L(L+1)j_{L}(k|R_1|)h^{+}_{L}(k|R_1|) \nonumber \\
& \cdot \langle a_{L^\prime}^{i1}(\omega)^{\dagger} \cdot A^{i1,i2}_{L^{\prime} L}
\cdot p^{i2}_{L}(\omega^{\prime}) \rangle ,\nonumber \\
\int_{S^2}\langle \mathbf{B}^{i1,out}(x;\omega)^{\dagger}\cdot
\mathbf{B}^{i2,in}(x;\omega^{\prime})\rangle|_{x=R[1]}
=&\Im 
\frac{k^2}{\omega^2}(L(L+1))j_{L}(k|R_1|)h^{+}_{L}(k|R_1|) \nonumber \\
& \cdot \langle a_{L^\prime}^{i1}(\omega)^{\dagger} \cdot A^{i1,i2}_{L^{\prime} L}
\cdot p^{i2}_{L}(\omega^{\prime}) \rangle .\nonumber
\end{eqnarray}
Here we have used the two relations that the eigenfunctions in the
bulk background satisfy:- 
\begin{eqnarray*}
\mathbf{L}|L,m,j_{L}(k|x|) \rangle= j_{L}(k|x|)\mathbf{L}|L,m \rangle, \\
\left(\frac{1}{k}\nabla \wedge \mathbf{L}\right)_r|L,m,j_{L}(k|x|)
\rangle
=\frac{L(L+1)j_{L}(k|x|)}{k|x|}|L,m \rangle .
\end{eqnarray*}
In addition we are anticipating a $\delta$-function in the frequency
so that we have been a little cavalier with the frequencies above.  

\subsection{Evaluating the background vacuum}

What remains to be calculated are the expectation values of the p-p
vacuum modes, $\langle
0|(p^{i}_{L1,m1})(p^{i}_{L2,m2})^{\dagger}|0\rangle$ with respect to
some sphere $i$ (in our case this will be sphere-1). This can
be evaluated for a dielectric background $\epsilon_B (\omega)$ that is independent of position and when there are no spheres present. We can then perform the integral over all
space. Note this is different from the case of two static dipoles
interacting as it is only the location of the dipole source where the
noise current is non-zero as opposed to here where the entire
background is filled with noise which the spheres polarizability
couples to. 

The composition rule for two free (or bulk) Green's functions in the
absence of any perturbing spheres integrated over all space is given
by (a more general result using functional differentiation and a
proper tensor treatment is given in~\ref{app:composition}) 
\begin{equation}
\int d^3z G^{free}(x,z|k)(G^{free}(y,z|k))^{\dagger}=\frac{1}{2k}
\partial_kG^{free}(x,y|k),
\end{equation}
with $k=\sqrt{\epsilon_B(\omega)} \omega/c$. Using the explicit form of the free propagator for incoming modes
\begin{equation}
G^{free}_{ij}(x,y|k)=(\delta_{ij}+\frac{1}{k^2}\nabla_i\nabla_j)
\frac{e^{i k |x-y|}}{4\pi |x-y|},
\end{equation}
which becomes
\begin{equation}
-i\partial_k G^{free}_{ij}(x,y|k)=\frac{1}{4\pi}(\delta_{ij}
+\frac{1}{k^2}\nabla_i\nabla_j)e^{i k |x-y|}.
\end{equation}
This can then be decomposed as a spherical wave expansion (see e.g.~\cite{Butler1999})
\begin{eqnarray}
-i\partial_k G^{free}_{ij}(x,y|k)&=& 4\pi (\delta_{ij}
+\frac{1}{k^2}\nabla_i\nabla_j) \nonumber \\
 & &\cdot \sum j_{L1}(k|x|)j_{L2}(k|y|)Y_{L1,m1}Y_{L2,m2}^{\dagger}. \nonumber
\end{eqnarray}
We can use this result to evaluate the background media vacuum that
has the quantum noise fluctuations throughout. The in field modes are
given by (at some fixed radial distance and in $\Sigma_{i}$ coordinate
system) 
\begin{equation}
p^{i}_{L1,m1}(\omega)=\frac{\omega}{j_{L1}(|x|)} \int_{S^2} \int d^3z
 \mathbf{L}Y_{L1,m1}^{\dagger}\cdot G^{free}(x,z|k)\cdot
\mathbf{J}(z;\omega). 
\end{equation}
The quantum noise current correlation functions at zero temperature are (see~\cite{raabe:013814})
\begin{equation} 
 \langle \mathbf{J}^{\dagger}_i(x,\omega)
\mathbf{J}_j(y,\omega^{\prime}) \rangle=0, 
\end{equation}
and
\begin{equation} 
 \fl\langle \mathbf{J}_i(x,\omega)
\mathbf{J}^{\dagger}_j(y,\omega^{\prime}) \rangle=\frac{\hbar }{\pi} \left(\frac{\omega^2}{c^2}\sqrt{\Im(\epsilon_B(\omega))}\sqrt{\Im(\epsilon_B(\omega^{\prime}))}\right)
\delta^3(x-y)\delta(\omega - \omega^{\prime}).
\end{equation}
After some algebra
and taking the limit $x\rightarrow y$ we obtain for the modes incident
on the $i$-th sphere 
\begin{equation}
\langle 0|(p^{i}_{L1,m1})(p^{i}_{L2,m2})^{\dagger}|0\rangle
= \frac{\hbar \omega^2}{2}\Im (ik)\delta(\omega -
\omega^{\prime}) \mathbf{1}^{i}_{L1,m1}\otimes \mathbf{1}^{i}_{L2,m2} ,
\end{equation}
where $\mathbf{1}$ is the $(2L+1)$ vector populated entirely by 1's
$\mathbf{1}=(1,\cdots ,1)$, which we shall from now on denote as
$|\mathbf{1}\rangle$. We see that there are no correlations between
the different mode numbers as one would expect from random noise. By
rotating to the imaginary frequency axis it is found to be a sensible
positive definite object representing the different occupation numbers
of the noise excitations. Further, the size of these vectors give us
the level at which we are truncating our series representation in the
angular momentum basis. 

\subsection{Perturbative force for dielectrics}

By making the appropriate substitutions from the previous sections,
one is able to calculate explicitly the force on sphere-1 due to
simply connected scattering processes between the $N$-spheres
(multiple reflections can be included in an obvious fashion). This is
given in~\ref{app:forcederivation}. Here we state simply the
result: 
\begin{eqnarray}
\label{eq:NSforceGeneral}
\fl \mathbf{F}[1|N-1]=-(-1)^N\frac{\hbar c}{4\pi}\sum^{N}_{i=2}
\nabla_{\mathbf{r}[i,1]} 
\int_{0}^{\infty}dX \frac{e^{-X}} {\sqrt{\epsilon_B}\mathcal{D}[1,1]}\cot \left(
\frac{\hbar c X}{k_BT\mathcal{D}[1,1]}\right)\times \nonumber \\ 
\langle\mathbf{1}|\left[ \mathcal{PO}[\alpha,\mathcal{A}]^N_{1,1}\right]  
\left[\frac{XR[1]}{\mathcal{D}[1,1]} j\left(\frac{XR[1]}{\mathcal{D}[1,1]}\right) n
\left(\frac{XR[1]}{\mathcal{D}[1,1]}\right) W
\left(\frac{XR[1]}{\mathcal{D}[1,1]}\right)\right]|
\mathbf{1}\rangle. 
\end{eqnarray}
Here, $\mathcal{D}[1,1]$ is the loop distance of the simply-connected
path from sphere-$1$ back to sphere-$1$ consisting of the sum of the $N$ separations, 
\[
\mathcal{D}[1,1]=\left( |\mathbf{r}[1,i]|+
|\mathbf{r}[i,j]|+\cdots |\mathbf{r}[j,k]|+|\mathbf{r}[k,1]|\right)_{s.c.}
\]
and $\mathcal{A}^{i+1,i}$ are the set of polynomials in
$(X,\mathbf{r}[i+1,i])$ that remain after extracting the exponentials
from the translation coefficients. Due to the additive nature of these
exponentials one finds it is the \emph{total path length} that occurs
in the exponential. The function $W$ is a measure due to the two
separate contributions in the stress tensor arising from the
individual field components and isotropic part (see~\ref{app:forcederivation}). 

This expression can be used as a perturbative expansion. The
arguments of the Mie scattering coefficients are always small since the total
path length will always be larger than the sphere radii. It can
then be evaluated in different perturbative setups when $T=0$, or by
residues when $T\neq 0$. It is also worth commenting how the expression for the force derived here will differ or coincide with that found from the Minkowski stress tensor or from energy functional methods. At least in some perturbation scheme there will be two main differences. One will be the scale of the force due to the differing factor of the background permittivity tensor. The second will be a different set curvature corrected terms that result from the extra isotropic contribution to the stress tensor. In the limit where the background permittivity is set to the vacuum value, all the expressions for the force will coincide in an exactly analogous way to that of the parallel plate geometry found in~\cite{raabe:013814}.

\section{The Force Between Two Spheres}
\label{sec:2sphereForce}
We now analyse the specific situation of the force on one sphere due
to multiple scattering with second in the retarded limit. In this manner we have an illustrative scheme in which to evaluate the force and thereby derive simple expansions. This was performed recently using functional determinants in~\cite{emig-2007-99,emig-2008-41} where in particular the scattering coefficients are expanded to higher powers in the frequency (and thereby sphere radii). For simplicity we consider
identical spheres that are aligned along the z-axis resulting in the
diagonalisation of the $m$-indices and take the static values of the different permittivities (since we are in the retarded limit). The Mie scattering coefficients are also expanded for small arguments only to first order so that we will find static polarizabilities of the simple multi-poles entering expressions. This is done to show the applicability of the method. All evaluations (integration and
matrix operations) have been performed in Maple and numerical data has been taken from~\cite{data}. 

\subsection{The retarded limit in the vacuum at $T=0$}

\begin{figure}[htbp]
\begin{center}
\includegraphics[height=8.0cm]{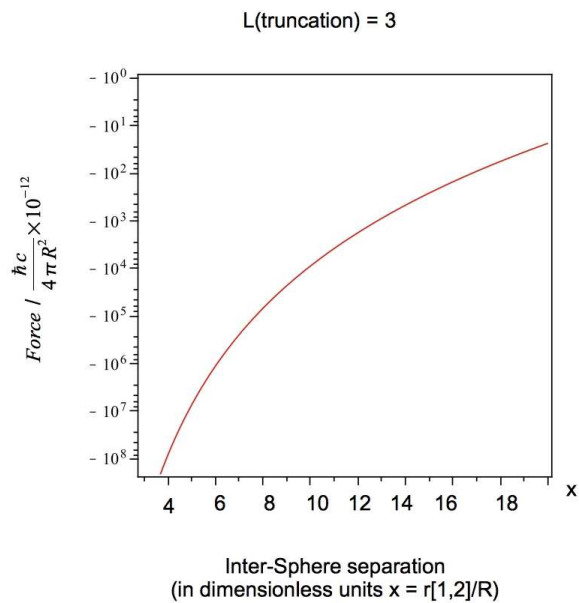}
\caption{A plot of the inter-sphere retarded force between two identical
dielectric spheres (with radii $R\equiv R[1]$) in the empty vacuum at $T=0^{\circ} K$. The vertical axis is the dimensionless ratio $\mathrm{Force}\:/\:(\hbar c/(4\pi R^2)\times 10^{-12})$, whilst the horizontal axis is the dimensionless ratio $x=r[1,2]/R$. Here, the
relative dielectric permittivity of the polystyrene spheres is
$\epsilon_1=\epsilon_2=2.6$ . The angular momentum series
representation has been truncated at
$L_{truncate}=3$.}\label{fig:TWOSPHERES} 
\end{center}
\end{figure}

For the case of two spheres in the vacuum, one is able to perform an
expansion in either the curvature of the spheres or the difference in
permittivities (or the product combination thereof occurring in the
scattering coefficients). In addition the $number$ of scattering
events also acts as perturbation parameter. The first perturbation
series we consider is the difference in permittivity to make contact
with known results. These can be assembled into the standard multipole
moments plus higher curvature corrected contributions that are valid
results at bilinear order in the difference in the
permittivities. This second set of terms arise from the isotropic part
of the stress tensor. Indeed, an expression is obtained directly for
the two body force 

\begin{equation}
\label{eq:multipoleForce}
F_{z}[1|2]=-\frac{\hbar c}{4\pi}\sum_{m,n=1}^{\infty}
\frac{1}{r[1,2]^{4+2m+2n}}
\left(v_{m,n}\alpha^1_m \alpha^2_n +
\frac{w_{m,n}\alpha^1_m \alpha^2_nR[1]^2}{r[1,2]^2} \right). 
\end{equation}
Here the coefficients $(v_{m,n},w_{m,n})$ are a set of numerical
coefficients found after evaluating the frequency integral, whilst the
multipole moments $\alpha^i_L$ are the multipole moments of sphere-$i$
given by 
\begin{equation}
\alpha^i_L=\frac{\epsilon^i/ \epsilon_B-1}{\epsilon^i/
\epsilon_B+(L+1)/L}R[i]^{2L+1}. 
\end{equation}
It should be said that the first term in the series coincides with the well known vacuum result with $v_{1,1}=23$. This has been evaluated for polystyrene balls in the vacuum with values given in Table~\ref{tab:vacuum}. The force has been plotted
against separation in Figure~\ref{fig:TWOSPHERES}. 
\begin{table}[htdp]
\caption{Material properties of the two sphere system. The dielectric
spheres are polystyrene balls in a background of the vacuum at
$T=0$K.} \label{tab:vacuum} 
\begin{center}
\begin{tabular}{|l|c|c|} \hline
Properties   & Sphere 1 & Sphere 2 \\ \hline
Permittivity (Sphere)  & 2.6 & 2.6\\ \hline
Permittivity (Background)  & 1.0 & 1.0 \\ \hline
Radii (m)  &$1.0\times 10^{-6}$ & $1.0 \times 10^{-6}$ \\ \hline
\end{tabular}
\end{center}
\end{table}%

\subsection{Finite temperature $T> 0$}

\begin{figure}[htbp]
\begin{center}
\includegraphics[height=8.0cm]{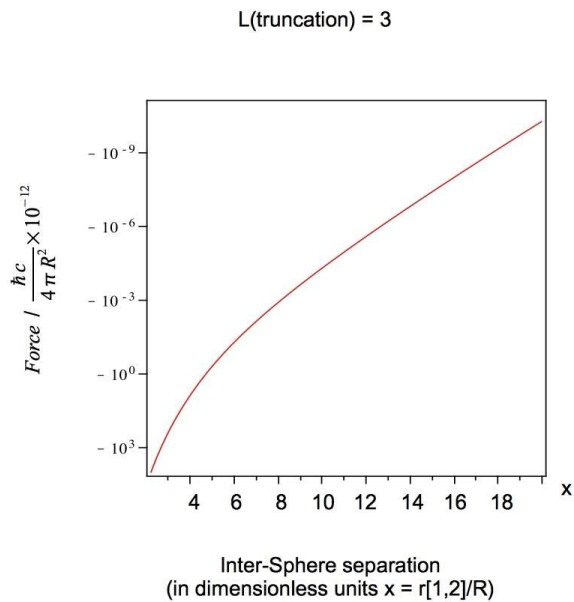}
\caption{A plot of the inter-sphere force between two identical dielectric
spheres (with radii $R\equiv R[1]$) in a silicone fluid background ($\epsilon_B=2.2$) at
$T=293^{\circ} K$. The vertical axis is the dimensionless ratio $\mathrm{Force}\:/\:(\hbar c/(4\pi R^2)\times 10^{-12})$, whilst the horizontal axis is the dimensionless ratio $x=r[1,2]/R$. Here, the relative dielectric permittivity of the
polystyrene spheres is $\epsilon_1=\epsilon_2=2.6$ . The angular
momentum series representation has been truncated at $L_{truncate}=3$
and the Matsubara frequencies truncated at
$l=2$.}\label{fig:2SpheresFiniteT} 
\end{center}
\end{figure}

Here we consider the polystyrene balls suspended in a liquid silicone
solution at room temperature. The values are given in
Table~\ref{tab:MATERIAL2}, followed by their plot in
Figure~\ref{fig:2SpheresFiniteT}. The poles are located on the
imaginary frequency axis at $\hbar cX_l/(k_BT \mathcal{D}[1,1])=l\pi$
for integer $l$. The attractive force can hence be evaluated to be of the power series form
\begin{eqnarray}
\label{eq:multipoleForceT}
F_{z}[1|2]=&-\frac{\hbar c}{4\pi}\left(\frac{k_BT}{\hbar c}\right)
\sum_{l=1}^{\infty}\sum^{\infty}_{m,n=1}
e^{-2r[1,2]\sqrt{\epsilon_B} K_BT l\pi/(\hbar c)}\nonumber \\ 
&\times \alpha^1_m(l) \alpha^2_n(l) \sum^{2m+2n}_{t=0}
\left( \frac{k_BT \pi l }{\hbar c} \right)^{2m+2n-t}
\left(\frac{g_t}{r[1,2]^{3+t}}\right)\nonumber \\ 
& \times \left(v^{\prime}_{m,n}(l)+w^{\prime}_{m,n}(l)
\left(\frac{k_BTR[1]}{\hbar c}\right)^2 \right).
\end{eqnarray}
Once again the integration of the frequency gives another set of
multipole type coefficients $(v^{\prime}_{m,n},w^{\prime}_{m,n})$
together with another set $g_t$ that arise due to the different
angular momentum states that constitute the translation matrices. 

\begin{table}[htdp]
\caption{Material properties of the two sphere system. The dielectric
spheres are polystyrene balls in a background of liquid silicone at
$T=290$K.} \label{tab:MATERIAL2} 
\begin{center}
\begin{tabular}{|l|c|c|} \hline
Properties   & Sphere 1 & Sphere 2 \\ \hline
Permittivity (Sphere)  & 2.6 & 2.6\\ \hline
Permittivity (Background)  & 2.2 & 2.2 \\ \hline
Radii (m)  &$1.0\times 10^{-6}$ & $1.0 \times 10^{-6}$ \\ \hline
\end{tabular}
\end{center}
\end{table}%

\section{The Force Between Three Spheres}
\label{sec:3sphereForce}

We now analyse the specific situation of the force on one sphere due
to multiple scattering with two other spheres, again in the retarded limit for illustrative purposes. See also~\cite{axilrod43, buhmann2006} for details of three atom potentials. It is worth commenting
that in order for the simply connected diagrams to hold as valid, we
require a clear line of sight  between the three spheres. This means
in particular that the multipole  expansion will not be accurate when
the three spheres are collinear  (indeed to form a connected diagram
now requires multiple internal reflections effectively convoluting two
diagrams that are both two-body). However, as with the approximations
made for the translation coefficients (centre-to-centre neglecting the
material properties of the spheres), we neglect this effect and assume
this collinearity effect not to be present. As with the two sphere case, static permittivities are used (corresponding to the retarded limit) and the Mie scattering coefficients are expanded for small arguments so that expressions contain static polarizabilities.

\begin{figure}[htbp]
\begin{center}
\includegraphics[height=7cm]{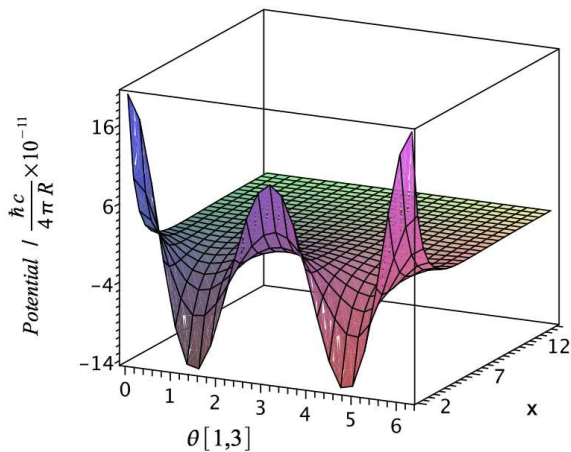}
\caption{A plot of the inter-sphere retarded potential  between three identical dielectric spheres (with radii $R\equiv R[1]$) in the empty vacuum
at $T=0^{\circ}K$. The vertical axis is the dimensionless ratio $\mathrm{Potential}\:/\:(\hbar c/(4\pi R)\times 10^{-11})$, the horizontal axis labelled by $x$ is the dimensionless ratio $x=r[1,3]/R$, and the remaining axis is the angular variable $\theta [1,3]$. Here, the relative dielectric permittivity of the polystyrene
spheres is $\epsilon_1=\epsilon_2=\epsilon_3=2.6$. Spheres 1 and 2 are
held fixed along the z axis $10R$ apart. The angular momentum
series representation has been truncated at
$L_{truncate}=1$.}\label{fig:THREESPHERES1} 
\end{center}
\end{figure}

\begin{figure}[htbp]
\centering
\begin{center}
\includegraphics[height=7cm]{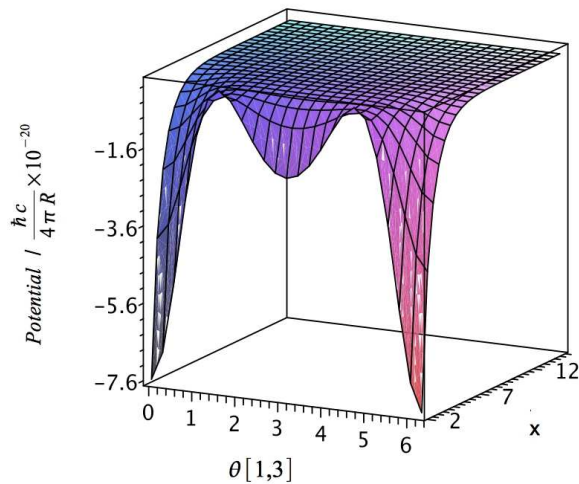}
\caption{A plot of the inter-sphere retarded potential between three identical dielectric spheres (with radii $R\equiv R[1]$) in silicone fluid
with $\epsilon_B=2.2$ at $T=293^{\circ}K$. Spheres 1 and 2 are held
fixed along the z axis $10 R$ apart. The vertical axis is the dimensionless ratio $\mathrm{Potential}\:/\:(\hbar c/(4\pi R)\times 10^{-20})$, the horizontal axis labelled by $x$ is the dimensionless ratio $x=r[1,3]/R$, and the remaining axis is the angular variable $\theta [1,3]$. Here, the relative dielectric
permittivity of the polystyrene spheres is
$\epsilon_1=\epsilon_2=\epsilon_3=2.6$ . The angular momentum series
representation has been truncated at
$L_{truncate}=1$.}\label{fig:C3SFT} 
\end{center}
\end{figure}

Figure~\ref{fig:THREESPHERES1} shows a plot of the inter-sphere
retarded potential from which the force can be found from simple
gradients given in Equation~(\ref{eq:NSforceGeneral}). This is given as
a function of the separation and angle for the third sphere for a
fixed separation vector between spheres 1 and 2 in the retarded limit
in the vacuum at $T=0$.  For the finite temperature case $T> 0$, Figure~\ref{fig:C3SFT} is a plot of the three sphere potential for a fixed separation of sphere 1 and 2 along the z-axis. 

\section{The Large $N$ limit of the Inter-Sphere Force}
\label{sec:Nsphere} 

Consider the case of $N$ identical spheres arranged in some random fashion as
$N\rightarrow \infty$, $\alpha^i =\alpha\rightarrow 0$, whilst
$\lambda :=N\alpha_S$ is held fixed ($\alpha=\alpha_S\omega^3R^3/c^3$). We put one constraint on the separation of all the spheres, that every sphere has two nearest neighbour spheres with a separation $s$ between them. In this case there is a single irreducibly connected diagram that has a minimum path length 
$\mathcal{D}[1,1]_m$. There are also approximately $N$ irreducibly connected diagrams that have a slightly increased path length $\mathcal{D}[1,1]_{m+1}$ that satisfy $\mathcal{D}[1,1]_m<\mathcal{D}[1,1]_{m+1} <\mathcal{D}[1,1]_m +2s$. We make the further approximation that each term of the second set contribute the same amount as the minimum path length diagram to the $N$-body scattering. The
expression for the associated potential reduces to 
\begin{eqnarray}
\label{eq:LARGENSPHEREFORCE2}
V[N]\approx \pm &(-1)^N\frac{\hbar c}{\mathcal{D}[1,1]}_m\times N \nonumber \\ 
& \times \int_{0}^{\infty}dX e^{-X}\cot
\left( \frac{\hbar c X}{k_BT\mathcal{D}[1,1]}_m\right)\left[  \alpha
\bar{\mathcal{A}}^(X,s)\right]^N. 
\end{eqnarray}
Taking the $T\rightarrow 0$  limit and then $N \rightarrow \infty$ we
can extract the dominant $L=1$ term  
\begin{equation}
\label{eq:LARGENSPHEREFORCE3}
V[N] \approx \pm (-1)^N \frac{\hbar c}{\pi s} \int_{0}^{\infty}dX  e^{-X}\left[\left(
\frac{\epsilon_S-1}{\epsilon_S+2}\right)
\frac{R^3}{s^3}+\mathcal{O}(X/N)  \right]^N, 
\end{equation}
which reduces to (using Sterling's approximation)
\begin{equation}
V[N]\approx \pm (-1)^N\hbar c
\frac{e^{-N}}{N!}\lambda^N\frac{R^{3N}}{s^{1+3N}}.
\end{equation}
The use of this formula could be in adding an additional sphere to the
system and measuring the resulting oscillation of the force compared
to the absolute force before the sphere is added. Alternatively it could serve to quantify the errors or to rearrange the perturbation series.

\section{Conclusions}
\label{sec:conclusions}

We have calculated Casimir forces between spheres using a multiple
scattering approach. This has been done at both zero and
finite temperature. In the case of the simply-connected three-body force, we have pointed out the need for non-collinearity, but that it is perturbatively not necessary. The total closed path length being always greater than the respective length scales of the
associated radii helps in quantifying a perturbative evaluation, for example the frequencies (found from the radii and total path length) below which the Mie scattering coefficients can be expanded. This is simply due to the fact that the propagator connecting all scattering centres in this simple setup contains an exponential factor of the total path length.  Additionally it addresses the technical problem of going beyond the proximity force approximation (PFA)~\cite{messina-2009-80} since, although continuity equations are applied at each material surface, there exists a solution of the field equations in each material medium. In general there will be both surface modes and bulk excitations in the bodies and calculations of Casimir forces should contain both~\cite{Henkel2009}. The multipole expansions found when the Mie scattering coefficients are expanded are also not singular when the spheres touch, since the sum of the two radii act as a physical lower cutoff (non-overlapping potentials).

For large $N$ and weakly scattering spheres, we have deduced
a scaling type formula for the simply-connected force terms based on a fixed separation set of
properties for the configuration. This could be useful for forces
where we want to understand the behaviour as a function of $N$ rather
than the detailed configuration (separations and orientations). It can also serve to quantify the errors in a multiple scattering expansion when truncating the series at this order. In a
subsequent paper one of us will develop the idea of the total path length
playing a key role in Casimir interactions. 

Theoretically and experimentally there are many open issues that need to be addressed in conjunction and naturally follow on. It is clear that substantial progress has been made both theoretically and experimentally in refining the accuracy of Casimir force calculations and measurements~\cite{RevModPhys.81.1827}. However, it should be pointed out that the theoretical calculations are for idealised geometries that do not necessarily reflect the experimental reality. One is faced with a potential disparity between the geometry encountered in the experiment, and those computable theoretically. We may for instance start off with a body that is very spherical, but that when we process it (heating, gluing, or coating) the sphericity of it will be reduced.  For instance in~\cite{zwol.041605} the initial borosilicate spheres are extremely smooth (roughness $<0.1\%$) but that the process of heating can induce surface roughness. And even if we calculate scattering coefficients for a given surface roughness profile, this is not necessarily what will be found experimentally. The problem requires both theoretical and experimental studies that are developed together. It is clear that we must perform a matching of experimental data (classical scattering data from the body) with calculations that allow for these unknown 'to be measured' parameters (for example non-spherical surfaces). Only with this interplay can one be confident that the real bodies are being quantitatively described and to what extent perfect geometries are a good approximation. It could be labelled under the 'realistic geometries' motif. For bodies in dispersive media, more needs to be understood on different material choices together with improvements experimentally, to be able to resolve the detailed difference between different proposals for physical observables. 

\ack
J.~B. wishes to thank Stefan Buhmann, Alex Crosse, and Rachele Fermani
for numerous helpful and constructive discussions whilst at Imperial College London, and John Gracey for
helpful comments. This work was supported by the SCALA programme of the European commission and the UK Engineering and Physical Sciences Research Council (EPSRC).

\appendix

\section{Composition of Green's tensors}\label{app:composition}

The bulk Green's tensor is defined by
\begin{equation}
(\nabla_{ij}+k^2_{ij}(x,\omega))G^{free}_{ij}(x,y|k)
=\delta_{ij}\delta^3(x-y),
\end{equation}
which can be rewritten as
\begin{equation}
G^{free}_{ij}(x,y|k)=\langle x|  \frac{1}{\nabla_{ij}+k^2_{ij}}|y\rangle.
\end{equation}
This can be functionally differentiated with respect to $k_{ij}$
\begin{equation}
\int k^2_{jm}\frac{\delta}{\delta k^2_{jm}}G^{free}_{in}(x,y|k)
=-\langle x|  \frac{1}{(\nabla_{ij}+k^2_{ij})}k^2_{jm} \nonumber \\
 \cdot \frac{1}{(\nabla_{mn}+k^2_{mn})}|y\rangle.
\end{equation}
A complete set of states can now be inserted and together with taking
the imaginary part we find 
\begin{equation}
\int \Im( k^2_{mn})\frac{\delta}{\delta k^2_{mn}}G_{ij}(x,y|k)\\
=-\int d^3z G_{i m}(x,z|k)\Im (k^2_{mn})(G_{nj}(y,z|k))^{\dagger}.
\end{equation}
At this point further simplification is possible when a known Green's
tensor is available together with the corresponding dispersion
relation. 

\section{Derivation of the multi-sphere force}
\label{app:forcederivation}
Substituting the scattering coefficients and the driving mode vacuum
expectation values into the  field correlation functions one finds the
perturbative force on sphere 1 due to multiple scattering is, 
\begin{eqnarray}
\label{eq:NSPHEREFORCE1}
\mathbf{F}[1|N-1]=&-(-1)^N\frac{\hbar }{4\pi } R[1]
 \Im \int_{0}^{\infty}d\omega k \nonumber \\ 
& \times \langle\mathbf{1}| [ \alpha^{1}(\omega R[1])
\sum_{i=2}^{N}A^{1,i}(\mathbf{r}[1,i]) \cdot \alpha^{i}( \omega R[i] )\cdots \nonumber \\
&\cdots \sum_{j=2}^{N}A^{i,j}(\mathbf{r}[i,j]) \cdot \alpha^{j} \sum^N_{j=2}\nabla_{\mathbf{r}[j,1]} A^{j,1}(\mathbf{r}[j,1])] \nonumber\\
& \times j(kR[1])
h^{+}(kR[1])W(\omega R[1]) |\mathbf{1}\rangle, 
\end{eqnarray}
where (here the background permittivity is $\epsilon_B$)
\begin{equation}
W(\omega R[1])=L(L+1)\left(L(L+1)-\frac{1}{2}(1+\epsilon_B)
\omega^2R[1]^2\right). 
\end{equation}
Here we have omitted the $SO(3)$ indices as in the earlier text, but
for the measure $W$ above, have stated how it depends on the labels.  This encapsulates all contributions of the force as an order $N$ polynomial in the scattering coefficients.
We use the reality properties of the polarizabilities and translation
coefficients after Wick rotation together with the necessary thermal
factor for the finite temperature system to rewrite this expression. The polynomials in the translation coefficients can be seen to be a path ordered expression linking all the scattering centres in a closed loop starting and finishing on sphere 1. We define this polynomial to be $\mathcal{PO}[\alpha,A]^N_{1,1}$ and that it is simply connected (i.e. a scattering coefficient for any given sphere only appears once). With the gradient defined always to act on the last translation matrix, we obtain for the force, 
\begin{eqnarray}
\label{eq:NSPHEREFORCE2}
\mathbf{F}[1|N-1]=&\frac{(-1)^{N+1}\hbar }{4\pi }\sum^{N}_{i=2}
\nabla_{\mathbf{r}[i,1]}\int_{0}^{\infty}d\Omega  \cot
\left(\frac{\hbar \Omega}{k_BT}\right) \nonumber \\ 
& \times \langle\mathbf{1}|  \left[ \mathcal{PO}[\alpha,A]^N_{1,1} \right]
\nonumber \\
&\times (k R[1])\cdot j(R[1])\cdot n(R[1])W(\Omega R[1])
|\mathbf{1}\rangle. 
\end{eqnarray}
We now make a change of integration variables to $X=\sqrt{\epsilon_B}\Omega \mathcal{D}[1,1]/c $, which is dimensionless, together with the introduction of the Jacobian factor $J$ and the extraction of the exponentials from the translation coefficients, so that 
\begin{eqnarray}
\label{eq:LARGENSPHEREFORCE}
\mathbf{F}[1|N-1]=&-(-1)^N\frac{\hbar c}{4\pi }\sum^{N}_{i=2}
\nabla_{\mathbf{r}[i,1]} \nonumber \\
&\times \int_{0}^{\infty}dX J^{-1}
\frac{e^{-X}}{\mathcal{D}[1,1]}\cot \left( \frac{\hbar c
X}{k_BT\mathcal{D}[1,1]}\right) \nonumber \\ 
&\times \langle\mathbf{1}|  \left[ \mathcal{PO}[\alpha,\mathcal{A}]^N_{1,1}\right]
\nonumber \\
&\times
\left[\left(\frac{XR[1]}{\mathcal{D}[1,1]}\right)\cdot j\left(\frac{XR[1]}{\mathcal{D}[1,1]}\right)\cdot n\left(\frac{XR[1]}{\mathcal{D}[1,1]}\right)\cdot W\right]|\mathbf{1}\rangle, 
\end{eqnarray}
with
\begin{equation}
J:=\frac{dX}{d\Omega}.
\end{equation}
For the calculations in this paper we will always take this Jacobian
to be the simple case of a constant dielectric.



\end{document}